\documentclass[nonacm, sigconf]{acmart}

\copyrightyear{2025} 
\acmYear{2025} 
\setcopyright{rightsretained}
\acmConference[CHI EA '25]{Extended Abstracts of the CHI Conference on Human Factors in Computing Systems}{April 26-May 1, 2025}{Yokohama, Japan}
\acmBooktitle{Extended Abstracts of the CHI Conference on Human Factors in Computing Systems (CHI EA '25), April 26-May 1, 2025, Yokohama, Japan}\acmDOI{10.1145/3706599.3719816}
\acmISBN{979-8-4007-1395-8/2025/04}
\usepackage{acmart-taps}
\usepackage{amsmath,amsfonts}
  
\usepackage{subcaption}
\usepackage{graphicx}

\usepackage{enumitem}

\newcommand{\enquote}[1]{``#1''}

\begin{document}

\title[LLM-Powered Interactive Privacy Policy Assessment]{\enquote{You don’t need a university degree to comprehend data protection this way}: LLM-Powered Interactive \\ Privacy Policy Assessment}

\author{Vincent Freiberger}
\affiliation{%
  \institution{Leipzig University}
  \city{Leipzig}
  \country{Germany}
}
\affiliation{%
  \institution{Center for Scalable Data Analytics and Artificial Intelligence (ScaDS.AI)}
  \city{Leipzig}
  \country{Germany}
}

\email{freiberger@cs.uni-leipzig.de}

\author{Arthur Fleig}
\affiliation{%
  \institution{Leipzig University}
  \city{Leipzig}
  \country{Germany}
}
\affiliation{%
  \institution{Center for Scalable Data Analytics and Artificial Intelligence (ScaDS.AI)}
  \city{Leipzig}
  \country{Germany}
}
\email{arthur.fleig@uni-leipzig.de}

\author{Erik Buchmann}
\affiliation{%
  \institution{Leipzig University}
  \city{Leipzig}
  \country{Germany}
}
\affiliation{%
  \institution{Center for Scalable Data Analytics and Artificial Intelligence (ScaDS.AI)}
  \city{Leipzig}
  \country{Germany}
}
\email{buchmann@informatik.uni-leipzig.de}

\settopmatter{printacmref=false}

\begin{abstract}

Protecting online privacy requires users to engage with and comprehend website privacy policies, but many policies are difficult and tedious to read. 
We present the first qualitative user study on Large Language Model (LLM)-driven privacy policy assessment. 
To this end, we build and evaluate an LLM-based privacy policy assessment browser extension, which helps users understand the essence of a lengthy, complex privacy policy while browsing. 
The tool integrates a dashboard and an LLM chat. 
In our qualitative user study (N=22), we evaluate usability, understandability of the information our tool provides, and its impacts on awareness. 
While providing a comprehensible quick overview and a chat for in-depth discussion improves privacy awareness, users note issues with building trust in the tool.
From our insights, we derive important design implications to guide future policy analysis tools.

\end{abstract}

%    <concept>
%        <concept_id>10002978.10003029</concept_id>
%        <concept_desc>Security and privacy~Human and societal aspects of security and privacy</concept_desc>
%        <concept_significance>500</concept_significance>
%        </concept>
%    <concept>
%        <concept_id>10003120.10003121.10011748</concept_id>
%        <concept_desc>Human-centered computing~Empirical studies in HCI</concept_desc>
%        <concept_significance>300</concept_significance>
%        </concept>
%    <concept>
%        <concept_id>10003120.10003121.10003124.10010868</concept_id>
%        <concept_desc>Human-centered computing~Web-based interaction</concept_desc>
%        <concept_significance>100</concept_significance>
%        </concept>
%  </ccs2012>

% \ccsdesc[500]{Security and privacy~Human and societal aspects of security and privacy}
% \ccsdesc[300]{Human-centered computing~Empirical studies in HCI}
% \ccsdesc[300]{Human-centered computing~Web-based interaction}


\maketitle

\section{Introduction}
\label{sec:intro}

Almost every interaction with companies, online services, smart devices, etc.\ leaves trails of personal data. 
Companies leverage techniques like hyper-personalization, powered by Artificial Intelligence (AI) and Machine Learning (ML) with real-time data sources~\cite{jain2021hyper}, to create user profiles and enable micro-targeting~\cite{chouaki2022exploring}. 
This results in vast privacy risks, such as automated influence~\cite{benn2022s}, manipulation~\cite{Martin2022}, and potential security breaches. While companies invest in acquiring and analyzing their users' personal data, users often lack awareness of the associated privacy risks~\cite{gerber2019investigating} or have distorted perceptions of them~\cite{gerber2019johnny}. %
Privacy regulations such as the GDPR~\cite{eu2016regulation} force companies to communicate data management practices and users' rights in \textit{privacy policies}, thereby empowering users to make informed decisions about their personal information. However, evidence shows that companies focus on compliance, targeting lawyers instead of users~\cite{schaub2017designing}, so users rarely read privacy policies~\cite{obar2020biggest}.
Using LLMs to automatically assess privacy policies is a promising approach to solve this issue~\cite{hamid2023genaipabench,rodriguez2024large,woodring2024enhancing}.
Yet, no prior work evaluates their impact on the user's understandability and risk awareness. 
Therefore, we conducted a scenario-based qualitative user study. It involved 22 participants from diverse backgrounds, acquired via different channels. To conduct this study, we developed a Chrome extension, PRISMe (Privacy Risk Information Scanner for Me), which combines LLM-based automatic privacy policy assessment with:
(i) an interactive dashboard; and
(ii) a chat for open conversations with the LLM with 
(iii) customizable explanations and responses that adapt to the user's preferences for detail and complexity.
We focus on three research questions:

\aptLtoX{\begin{enumerate}%
	\item[\textbf{RQ.1}]\label{item:rq-understandable}
    How do users with varying privacy knowledge interpret PRISMe’s privacy policy explanations?
	\item[\textbf{RQ.2}]\label{item:rq-awareness} 
    How does using PRISMe shape users’ awareness of privacy risks? 
	\item[\textbf{RQ.3}]\label{item:rq-usability} 
    How suitable and usable is PRISMe for everyday use across different user contexts and tasks?
\end{enumerate}}{\begin{enumerate}[label=\textbf{RQ.\arabic*}]
    \item\label{item:rq-understandable}
    How do users with varying privacy knowledge interpret PRISMe’s privacy policy explanations?
    \item\label{item:rq-awareness} 
    How does using PRISMe shape users’ awareness of privacy risks?%
    \item\label{item:rq-usability} 
    How suitable and usable is PRISMe for everyday use across different user contexts and tasks?
\end{enumerate}}

Our findings suggests that a tool such as PRISMe can tremendously help users who lack awareness and comprehension regarding online-privacy risks~\cite{gerber2019johnny}, by communicating relevant privacy protection information. Moreover,
users tend to request evidence for the provided information (which PRISMe can provide). 

Our \textbf{main contribution} is a qualitative user study evaluating LLM-based privacy policy analysis. Our findings mark an important step toward developing interactive tools that enhance privacy awareness and informational sovereignty.

\section{Related Work}\label{sec:related}
\subsection{Challenges with Privacy Policies}
\label{sec:policy}

Privacy policies aim to mitigate the \emph{information asymmetry} between service providers and users~\cite{malgieri2020concept,Zaeem2020}. However, they are often designed for legal compliance, with dense, lawyer-centric language~\cite{schaub2017designing}. 
This makes it difficult for users to match them with their preferences~\cite{windl2022automating,mhaidli2023researchers}. Legal regulations, e.g., GDPR~\cite{eu2016regulation} or \cite{CaliforniaStateLegislature2018}, do not prevent persuasive language, which might obscure unethical practices and create a false sense of trust~\cite{pollach2005typology,belcheva2023understanding}. 
Thus, users reading and understanding privacy policies is rare~\cite{reidenberg2015disagreeable,steinfeld2016agree}, which results in informational unfairness~\cite{freiberger2024legal}.
Generative AI~\cite{lee2024deepfakes} and Augmented Reality complicate data management practices further and exacerbate transparency issues~\cite{becher2021law,belcheva2023understanding,transparency}.

\subsection{The Landscape of Transparency-Enhancing Technologies}\label{sec:assistants}

Early \emph{approaches reliant on a privacy language representation of the policy} like Privacy Bird~\cite{cranor2006user}, relied on the P3P privacy language~\cite{cranor2002web} to preconfigure privacy sensitivity levels and used static, rule-based risk assessments to display warnings via a traffic-light system. Current privacy nutrition labels~\cite{kelley2009nutrition} provide a similar visualization. However, fixed criteria limit adaptability, and the lack of user-query mechanisms hinder interactivity and education. Inspired by these, PRISMe evaluates policies in the background and presents results succinctly. Grünewald et al.~\cite{grunewald2023enabling} later introduced layered dashboards with chatbot interaction but with limited dialogue flexibility. We also present minimal information initially, allowing users to uncover additional details based on chat interactions.

Tools like Poli-see~\cite{guo2020poli} and ToS;DR~\cite{Roy2024} \emph{rely on processed data}, e.g., crowd-sourced annotations. Poli-see visualizes data flows via icons on dashboards, while ToS;DR uses color-coded summaries for quick evaluation. Although effective at simplifying complexity, reliance on static data limit its scalability and adaptability.

Advances in \emph{automated assessment with Natural Language Processing (NLP)} shifted focus to plain-text privacy policies. Polisis~\cite{harkous2018polisis} and PriBot~\cite{harkous2016pribots} utilized privacy-specific embeddings and ML classifiers to identify policy elements, allowing users to query content. Since the returned segments are direct policy quotes, users may still struggle with comprehension. In contrast, our LLM-based chat can deliver clearer explanations, cite policy evidence upon request, and handle a broader range of questions beyond the policy text.
PrivacyInjector~\cite{windl2022automating} enhanced decision-making through context-aware visualizations and explanations. While promising to improve privacy awareness, users suggested reducing text length and interpreting the severity of privacy threats, which PRISMe makes possible.
PrivacyCheck~\cite{nokhbeh2020privacycheck} evaluates privacy policies using 20 static questions on user control and GDPR, 17 of them yes/no questions, scored by ML models. It identifies the website's market sector and compares scores with three competitors. However, its fixed questions and binary responses limit user interaction and the ability to query or simplify explanations.

Tools based on \emph{ML classifiers}, e.g., Claudette~\cite{contissa2018claudette}, PrivacyGuide~\cite{tesfay2018privacyguide} and GDPR-completeness classifiers~\cite{torre2020ai,amaral2021ai,xiang2023policychecker} evaluate binary or check-list-based scores on GDPR compliance. However, they often fail to communicate implications effectively to non-lawyers and do not consider individual privacy preferences. %

\emph{LLM-based privacy policy assessment} has shown to be as effective as NLP methods in extracting key aspects from privacy policies, such as contact information and third parties~\cite{rodriguez2024large}.
ChatGPT-4 offers performance and adaptability in answering privacy-related questions~\cite{hamid2023genaipabench}, which motivated our use of GPT-4o~\cite{OpenAI2024}. 
Privacify~\cite{woodring2024enhancing} is a browser extension summarizing policy chunks for compliance analysis and data collection insights. However, it lacks customization, interactive features, and, most importantly, a comprehensive evaluation.

\begin{figure*}[htb]
    \centering
    \includegraphics[width=0.9\textwidth]{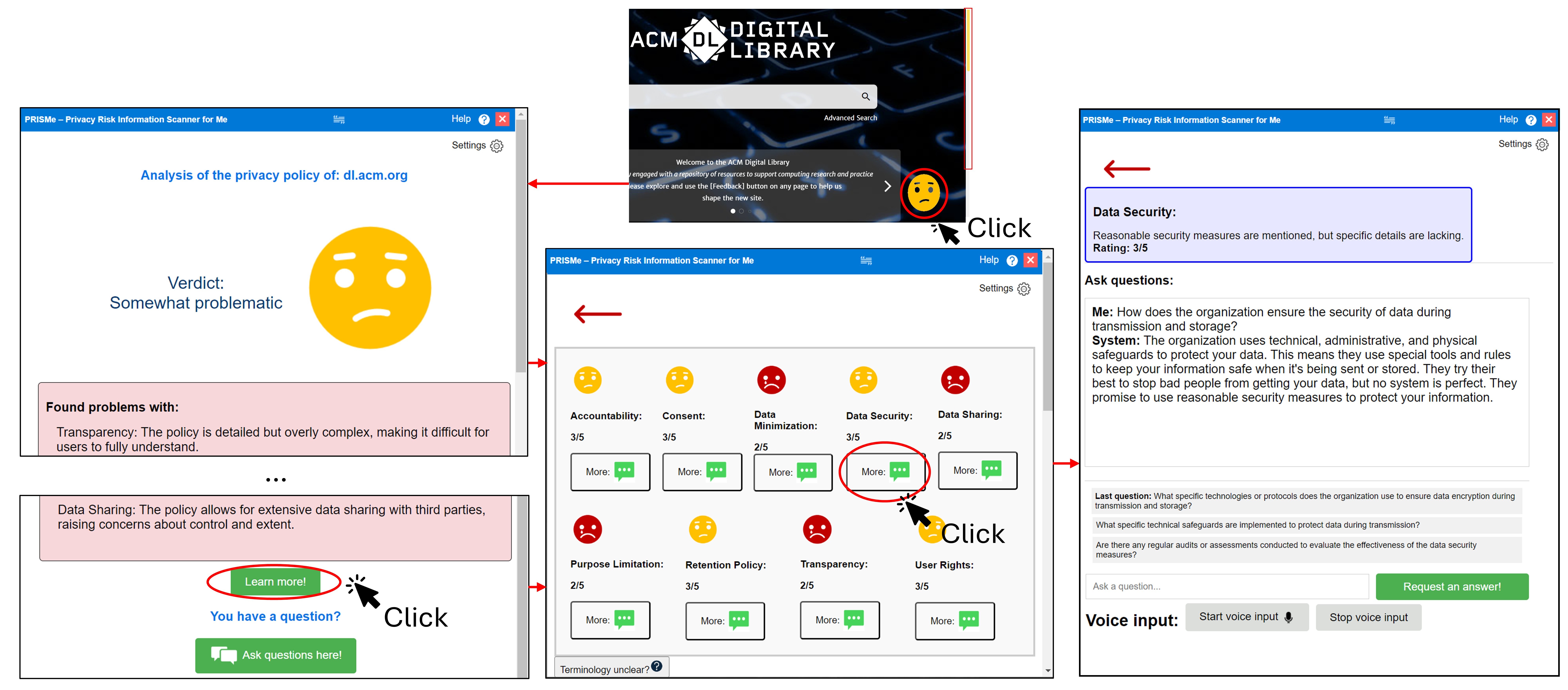}
    \caption{When the user visits a website, our prototype evaluates the privacy policy in the background and displays privacy alerts via colored scrollbars and a point-of-entry smiley icon (top middle). Clicking the smiley opens an Overview Panel (left) summarizing key privacy issues, with navigation to a Dynamic Dashboard and chat interface. The dashboard (bottom middle) provides detailed policy evaluation criteria, which users can chat about (right) by clicking the respective "More" button.}
    \label{fig:ui}
\end{figure*}

\subsection{Identified gaps for AI tools assessing privacy policies}
Since \textit{machine-readable privacy policy representations} are rare, tools like Privacy Bird offer inspiration but lack real-world applicability. Tools relying on \textit{processed crowd-sourced data} face scalability and adaptability challenges. \textit{Compliance-focused approaches}, while valuable for legal experts, struggle to address typical users. \textit{Static criteria} limit adaptability to users' needs and evolving practices, and the \textit{lack of interaction} in many tools hinders user engagement, awareness and comprehension. LLM-based tools remain nascent, with no qualitative user studies conducted.

\section{Study Methodology}
\label{sec:met}

We conducted an exploratory lab-based qualitative user study (N=22) using a research prototype of PRISMe in three scenarios. We outline our scenarios first (full user instructions in Appendix~\ref{sec:scenario}), and explain PRISMe in Section~\ref{sec:design-process}. Participants could spend as much time as they wanted in each scenario.

\textbf{Scenario 1 "Privacy Exploration on a News Media Platform (focus.de) and Payment Provider (PayPal)"} allows to assess how users engage with privacy policies using PRISMe, targeting~\ref{item:rq-understandable} \textit{(understanding)} and \ref{item:rq-awareness} \textit{(awareness)}.
It exposes participants to common yet complex services with significant data collection. 

\textbf{Scenario 2 "Comparing Privacy Practices"} addresses \ref{item:rq-usability} \textit{(usability)}. Using the setting of four German online bookstores (same pricing, different levels of data protection) before a hypothetical purchase, users are instructed to compare privacy policies and decide. This exposes users to a realistic, comparative setting where they choose between services. The presentation order of bookshops in the scenario description was changed between participants.

\textbf{Scenario 3 "Free Exploration of Websites"} examines how PRISMe supports users’ privacy concerns in personalized contexts. Participants freely explore websites of their choice, providing insights into the tool's real-life engagement potential, contributing to \ref{item:rq-understandable} \textit{(understanding)}, \ref{item:rq-awareness} \textit{(awareness)}, and \ref{item:rq-usability} \textit{(usability)}.

\subsection{Experimental Design and Procedure}
After signing a consent form, participants completed a questionnaire on demographics, privacy attitudes, and browsing habits.
Next, the facilitator demonstrated and explained PRISMe's functionalities. 
Afterwards, the participants explored the tool independently with suggested or self-chosen websites until they felt comfortable with PRISMe, with guidance available if needed. 
They then completed the three scenarios, averaging 29.8 minutes.%
The facilitator encouraged the participants to ask questions and voice comments, which were documented. After completing the scenarios, participants filled out the SUS questionnaire and our custom questions (see Figure~\ref{sec:questionnaire}). 
The study concluded with a semi-structured interview (see Appendix~\ref{sec:guide} for the interview guide) lasting an average of 17.6 minutes. 
The interviews were prepared and conducted according to Myers~\cite{myers2020qualitative}.
We transcribed the interviews using faster-whisper (large-v3)~\cite{systran2023fasterwhisper,openai2024whisperlargev3} and manually checked the transcripts for accuracy and consistency. With an \textbf{open coding} approach based on grounded theory~\cite{corbin1990grounded,corbin2015basics}, we analyzed the transcripts and comments made by participants during the scenarios facilitated by Taguette~\cite{remram44_taguette}. Coding was done by two researchers independently. After initial coding, both authors reviewed the other's codes before an in-person discussion addressing clarification (23 codes) and disagreements (15 codes) leading to 3 codes entirely dropped or aggregating 8 codes to higher levels of abstraction. It also resolved the slightly different naming of codes. The result is an aggregated and cleaned set of 61 codes (896 coded passages). Based on all codes, both authors identified overarching topics that codes belong to and grouped them accordingly before aggregating their results in another discussion leading to 6 topics (after each author identified 5 initially).

\subsection{Participants}
In two German cities (Leipzig and Chemnitz) we recruited 22 participants (14 male, 8 female; age range: 18-64) via mailing lists, online message boards, public events on AI, adult education centers, and convenience sampling. The participants were researchers in the fields IT (3), industrial production (3), chemistry (2), law (1), students at a university (4) or professionals in the areas IT (3), education (2), real estate (1), entertainment (1), crafts (1) and healthcare (1).
Our participants' expertise on data protection ranges from novices to experts, including a data protection officer. All of our participants rarely or never read privacy policies. 
Participants were compensated with a 15 Euro gift card. The study was scheduled for one hour, but participants were free to explore further websites of their choice. We measured durations from 60 to 90 minutes.

\section{Our PRISMe Prototype}\label{sec:design-process}

For the practical part of our study, we implemented PRISMe (see Figure~\ref{fig:ui}), a Chrome browser extension. 
Section~\ref{sec:related} revealed a general sense of \textit{helplessness}, \textit{low involvement with privacy-related decisions}, and a preference for a smooth \textit{browsing experience} instead. This led us to four \textbf{design considerations} for an LLM-based assessment tool for privacy policies: 

\aptLtoX{\begin{enumerate}%
	\item[\textbf{DC.1}]\label{item:dc-clear-communication} Communication should be clear, adaptable, and comprehensible for a wide range of users.
	\item[\textbf{DC.2}]\label{item:dc-nondisruptive} The tool should not disrupt the browsing experience too much and should offer immediate feedback.
	\item[\textbf{DC.3}]\label{item:dc-playful-exploration} There should be an exploratory, ideally easy and engaging aspect to understanding privacy policies.
	\item[\textbf{DC.4}]\label{item:dc-adaptability} The tool should adapt to different privacy requirements and across various types of websites.
\end{enumerate}}{\begin{enumerate}[label=\textbf{DC.\arabic*}]
     \item\label{item:dc-clear-communication} Communication should be clear, adaptable, and comprehensible for a wide range of users.
     \item\label{item:dc-nondisruptive} The tool should not disrupt the browsing experience too much and should offer immediate feedback.
     \item\label{item:dc-playful-exploration} There should be an exploratory, ideally easy and engaging aspect to understanding privacy policies.
     \item\label{item:dc-adaptability} The tool should adapt to different privacy requirements and across various types of websites.
\end{enumerate}}

After implementing a prototype, we tested it within our research group and refined it in iterative feedback cycles. Then we ran a pilot study with 4 male and 2 female participants different from the main study. Based on their feedback, we improved performance by caching LLM assessments and user input, and improved the scraping of policies. We re-arranged frontend interface elements and added a plain-text display of the website's policy. We also took into account individual preferences on policy assessment right from the start, and added a speech-to-text input method.

Figure~\ref{fig:ui} illustrates the PRISMe prototpye as used in our study.
When users switch tabs or visit a new website, PRISMe fetches and analyzes the privacy policy, highlighting concerns with color-coded scrollbars and smiley icons (Figure~\ref{fig:ui} top middle). The LLM (OpenAI's GPT-4o~\cite{OpenAI2024}) dynamically selects and evaluates criteria on a 5-point Likert scale with the system prompt provided in Appendix~\ref{app:prompting}. 

The smiley icon (green, yellow, red) summarizes the policy’s overall rating.\footnote{Based on the mean of the criteria scores (red: below 2.5, yellow: between 2.5 and 3, green: above 3). Thresholds were hand-calibrated. If the policy cannot be scraped, a gray question mark is displayed.}
Clicking the smiley opens the Overview Panel (Figure~\ref{fig:ui} left), summarizing critical issues (scores below 3) and linking to the dashboard and chat interfaces for deeper exploration.
The Dashboard Panel (Figure~\ref{fig:ui} bottom middle) provides detailed scores for each criterion, rated with smiley icons.\footnote{red: below 3, yellow: 3, green: above 3)} 
It also explains the score below the dashboard.
Users can explore criteria via a Criteria Chat (Figure~\ref{fig:ui} right) or use a General Chat (different system prompt) for broader questions. Both chats offer dynamically generated question suggestions (with GPT4o-mini~\cite{o_mini}), inspired by Ravichander et al.~\cite{Ravichander2021}, and allow typing or voice input.
Settings, accessible via a cogwheel icon, let users customize chat response length and policy assessments (short, medium, long) and complexity (beginner, basic, expert) to match their preferences and technical expertise.

\section{Results}
\label{sec:eval}

\subsection{Interview \& Comments Data}
We identified 61 distinct codes in 6 topic areas, which we summarize below (full codebook in Appendix~\ref{sec:codebook}). 

\textbf{Topic 1: User Attitudes, Motivations, and Behavior (\ref{item:rq-awareness}, \ref{item:rq-usability}).}
Participants often displayed \textit{indifference} (P1, P6, P8), \textit{insecurity} (P4, P14) or \textit{misconceptions} about privacy risks (e.g., believing private browsing prevents tracking) (P5, P9) and expressed distrust in websites' data protection practices (P1, P2, P8, P9, P11, P12).

However, PRISMe sparked emotional responses, curiosity, and increased \textit{privacy awareness}.
For instance, P1 stated, \textit{\enquote{a sad face like that does something to me emotionally}} and reflected on the importance of data protection, saying, \textit{\enquote{Looking at privacy policies has never been relevant in my life before, [...] but I think I need to become more aware}}. Participants were motivated to explore further, with P15 noting, \textit{\enquote{It was interesting, so I always wanted to try more}}. Participants were interested in using PRISMe in their daily lives: \textit{\enquote{If it was available, I think I would directly install it}} (P21), \textit{\enquote{I think it would be great progress if this became standard practice}} (P20). 

While participants see challenges in changing habits (P1, P4, P5, P6, P10, P18) and deem privacy-unrelated factors (more) important (P3, P8, P11, P12, P14, P16, P20, P22), they highlight professional applications, such as data protection training (P18), improving website practices (P20), documentation (P20), and assessing business partners (P6, P22).

\textbf{Topic 2: Information Quality and Clarity (\ref{item:rq-understandable}, \ref{item:rq-usability})}.
\label{sec:info_quality_interviews}
Participants praised PRISMe for \textit{simplifying} complex privacy policy language, acting as a "translator" (P20) that made data protection accessible, with P19 noting, \textit{\enquote{You don’t need a university degree to comprehend data protection this way}}. 
While some participants noted a lack of transparency in how ratings were determined (P1, P10, P12, P17, P18, P22) and expressed confusion over the rationale behind certain evaluations, \textit{\enquote{I couldn’t really understand why this lower rating was given. Or why the system came to this exact conclusion}} (P14), all agreed the simplified language significantly improved understanding.
All but two participants highlighted the quick, clear overview of privacy policies, enabling access to essential information with minimal effort. 

Feedback on the tool’s \textit{detail levels} was mixed. Some participants found discrepancies between overall ratings and alerts (P8, P11, P12, P18) or desired more actionable insights, such as keywords like "This is dangerous" (P10). Others acknowledge the balance between detail and usability, with P22 stating, \textit{\enquote{I would have liked it to be more specific, but then again you have to read more}}.
Participants found the information comprehensive and relevant. P2 summarized, \textit{\enquote{Everything important in a privacy policy is included: what data is collected, transparency, legal basis, purposes...}}.

The \textit{chat functionality} was well-received for its flexibility, handling typos, multiple languages, and sophisticated or nuanced queries (P1, P7, P15). Responses were consistent and conversational, often anticipating user needs (P9). As P22 noted, \textit{\enquote{I was surprised by how exact it answered with very specific information}}.

\textbf{Topic 3: User Experience and Interface Interaction (\ref{item:rq-usability}).}
\label{sec:experience_interviews}
Participants liked the smiley icons as \textit{visual cues} providing quick, intuitive insights. P9 stated, \textit{\enquote{The smiley is great because it doesn’t interfere but [quickly shows]: Is this [privacy policy] good or not?}}
The tool’s ease of use was widely praised, described as \textit{\enquote{user-friendly, clear, and courteous}} (P6). Some participants suggested improvements, like avoiding scroll bars (P8, P20), and reorganizing chat formatting for clarity (e.g., structured paragraphs and key highlights) (P1, P8, P10, P11, P13, P14, P18, P21).
Voice input was used and valued by some, and adding audio output for accessibility was suggested (P5, P12).

\textbf{Topic 4: Tool Reliability and Trustworthiness (\ref{item:rq-usability}).}
While participants generally appreciated the tool, concerns about reliability arose. P7 suspected  hallucinated or speculative information, while P8 questioned the source of the information: \textit{\enquote{Sometimes I doubted whether the given information was from the policy itself. It was sometimes expressed more like an assumption}}.
To enhance trust, participants suggested links to the relevant policy sections (P2, P8, P10, P13, P18, P22). Concerns were also raised about the dynamic evaluation criteria, which some felt lacked weighting by importance, reducing comparability between policies and perceived reliability (P3, P9, P12, P17, P22).

\textbf{Topic 5: Recommendations for Features and Functional Improvements (\ref{item:rq-usability}).}
Participants wished for a more solution-oriented design, like suggesting alternatives (P1, P4, P7, P12) and providing contextual feedback on current privacy settings (P15, P18). Some were open for allowing PRISMe to adjust browser privacy settings, potentially influencing the displayed ratings (P7, P8, P15).

\textbf{Topic 6: Impact on Users (\ref{item:rq-awareness}).} %
\label{sec:impact_interviews}
The tool generally raised participant's \textbf{privacy awareness}. 
P12 commented, \textit{\enquote{I think it helped me to be a bit more sensitive to the topic of data protection. I think this short session was already useful}}. 
Others reflected on typically ignoring privacy policies but expressed a new intent to pay closer attention (P1, P9). The tool also encouraged critical thinking about data protection. P11 mentioned, \textit{\enquote{I thought about what data protection problems could actually be.}}
Participants reported improved \textit{understanding} of relevant privacy issues, with P4 highlighting the value of explanatory sentences for evaluation criteria. This newfound awareness prompted some to consider being more cautious with their data: \textit{\enquote{I can imagine that I would be more careful about who I entrust with my data}} (P12). 

\subsection{Chat Responses}
We identified issues in 60 of 368 (16.3\%) responses of the LLM to chat queries, which we categorized into system-related and LLM-related problems outlined below. For numbers of occurrences and examples, see Appendix~\ref{sec:chat_res}.

\textit{Problems related to our system:} 
Four policies during free exploration (Scenario 3) were \textbf{partially scraped}, leading to 13 incomplete responses. The LLM avoided hallucinating details, offering abstract responses based on available content.
Four cases involved \textbf{context-related limitations} due to reliance on plain text rather than HTML. One missed a hyperlink to a requested form, and three failed to provide step-by-step privacy setting instructions.

\textit{Problems related to the LLM:} 
The LLM mirrored \textbf{overly euphenmistic}, marketing-like language from policies 12 times, adding reassuring but uncritical conclusions that irritated some participants (P9, P12, P17).
We found 12 cases where the LLM partially ignored or \textbf{omitted relevant information} in its chat responses, unless explicitly asked. This led to user frustration, particularly among study participants seeking specific details (P1, P3, P5, P17).
Eight \textbf{Generic responses} lacked specificity to the given policy and were perceived as less useful as they required additional prompts for clarification (P3, P9, P17).
Five responses included \textbf{hallucinations}, with one potentially harmful case recommending a less privacy-compliant service. Other hallucinations involved fabricated information about its own evaluation or audit details not specified in the policy.
\textbf{Misunderstandings of user queries} were evident in four responses. While these interrupted the flow of interaction, they were typically easy for users to identify and fix with follow-up queries (P8, P12, P16).
Two chat responses confused users by presenting seemingly contradictory information, sending \textbf{mixed signals}. Despite being technically correct, their phrasing caused unnecessary doubt or hesitation (P11, P13).

\section{Discussion}
\label{sec:disc}

\subsection{Behavioral patterns}

Based on how participants used PRISMe (e.g., number and complexity of questions asked) and interview responses, we distinguish participants into four categories with unique behavioral patterns and privacy needs. 
These profiles offer a framework for tailoring privacy tools to diverse user groups.

{\textit{Targeted Explorers}} (P2, P7, P12, P17, P18, P20, P22) engage deeply, seeking detailed and specific information. They tend to have prior privacy knowledge and clear goals, require advanced customization options, and request evidence.

\textit{Novice Explorers} (P4, P13, P14, P19) have limited prior knowledge and confidence in understanding privacy policies. Their exploration is more about discovering and defining their informational goals rather than coming in with a predefined agenda. Privacy tools for this group should emphasize simplicity and guidance, making learning accessible.

\textit{Balanced Explorers} (P1, P3, P8, P9, P10, P11, P15, P21) regularly combine discovery and seeking of specific information. They engage with the tool in a balanced manner benefiting from flexibility and broad overviews.

\textit{Minimalistic Users} (P5, P6, P16) prioritize efficiency, engaging minimally with the tool, are satisfied with high-level summaries and  usually stop after a few interactions. For this group, concise assessments and quick insights are essential.

\subsection{\ref{item:rq-understandable}: %
How do users with varying privacy knowledge interpret PRISMe’s privacy policy explanations?
}
A key insight is that our browser extension \textbf{simplifies complex privacy policy language}, which all participants -- even \textit{Targeted Explorers} -- identified as crucial in enhancing users’ understanding of privacy policies. 
\textit{Novice Explorers} in particular benefit from a "translator" from legalese to plain language.
These simplification capabilities are in line with research validating the text simplification potential of LLMs~\cite{kew-etal-2023-bless}. They successfully address the gap in previous transparency-enhancing tools to provide easily comprehensible explanations and interpretations for users, and not just policy quotes~\cite{harkous2018polisis,tesfay2018privacyguide,windl2022automating}.
The LLM also overlooks spelling mistakes and \textbf{adapts to different languages}, thus lowering entry barriers. Our voice input facilitates inclusion, and participants' appreciate \textbf{different modes} of interaction with such a tool. 
To enhance understanding, future work could improve the \textbf{response formatting}, e.g., clearer visual structuring and keyword highlighting. 
\textbf{Adjustable settings} span the full range from simple explanations for \textit{Novice Explorers} to  specific details in a technical terminology for \textit{Targeted Explorers}. 
\textbf{Personalization}, by breaking down complexity to a level at which the user can actively engage, was appreciated by users and should be a major goal for tools like ours. 
Personalization and high-quality suggestions are essential to avoid discriminatory outcomes, as the cognitive demand of asking the right questions or finding suitable settings may prefer high-literacy persons~\cite{Ravichander2021}. Existing tools only facilitate customization to privacy sensitivity levels~\cite{cranor2006user}, which mirrors our settings on output length, but they did not allow adjusting the complexity of explanations.
Potential next steps include \textbf{automated customization}, as discussed in~\cite{Ravichander2021,goram2023human}, and individual evaluation criteria, targeting decreasing metacognitive demands involved~\cite{tankelevitch2024metacognitive}.

\subsection{\ref{item:rq-awareness}: %
How does using PRISMe shape users’ awareness of privacy risks?} 
PRISMe increases \textbf{awareness} across all user profiles, 
with \textit{Novice Explorers} and \textit{Minimalistic Users} showing the most and \textit{Targeted Explorers} with pre-existing knowledge showing the least change. Our extension encourages \textbf{reflection} especially for \textit{Balanced Explorers}. 
It prompts them to engage with privacy issues more deeply by triggering \textbf{emotional responses}.
Future work might investigate to what extent our always-displayed initial smiley raises awareness. 

A potential risk is over-reliance on the tool’s assessments, which mitigates awareness by users feeling protected~\cite{schaub2016watching}.
The convincing nature of LLM outputs may mislead users into misjudgments~\cite{krugel2023chatgpt}. This is a risk that for instance PrivacyInjector~\cite{windl2022automating} does not face while still managing to raise user awareness.
Future tools should \textbf{manage expectations} and balance the persuasive nature of LLM outputs with mechanisms that encourage \textbf{critical thinking}, such as expressing uncertainty~\cite{kim2024m} or offering additional evidence.

\subsection{\ref{item:rq-usability}: How suitable and usable is PRISMe for everyday use across different user contexts and tasks?} 
Users of all groups find PRISMe easy and intuitive to use, as evidenced by a SUS rating of 88.9. 
They suggested improving details, such as an adjustable window size, the chat formatting, and confusion regarding the chat input and response. 
Most notably, participants require no support from a technical person. 
While we did give participants a hands-on tutorial before the study, this suggests that a \textbf{one-time tutorial} after installation should suffice. Windl et al. even rely on \enquote{serendipitous discovery} for user familiarization.

They are challenged however by evaluation criteria varying between websites and miss weighting of criteria by importance. Identifying and defining a set of fixed user-centric criteria may improve evaluation transparency and comparability. Weighting criteria is challenging as their importance is dependent on user preferences. Related tools also lack weighing their mostly GDPR-based evaluation criteria~\cite{tesfay2018privacyguide,nokhbeh2020privacycheck}.
Identifying baseline criteria weights combined with customization options or automatic detection of a user's preferences to change weighting require further research.

Participants appreciated the \textbf{layered design} and \textbf{customization options} across contexts, with some changing the settings during the study. 
This allowed them \textbf{fine-grained control} of the level of detail they explore. 
\textit{Novice Explorers} and \textit{Balanced Explorers} show the most interest in using the tool in their daily personal lives, with some describing their experience as playful.
Particularly \textit{Targeted Explorers} voiced a desire to use the tool in a professional setting.

\subsection{LLM limitations and further considerations}\label{subsec:llm-limitations}
Hallucinations are a limitation of LLM-based systems. In our analysis of 368 chat responses, hallucinations were rare (5 instances), typically harmless, and easily identifiable.
For example, the LLM once recommended a factually worse web service, a behavior potentially addressable by giving more general guidance.
\textit{Minimalistic Users} and \textit{Novice Explorers}, who prioritize efficiency or trust the tool uncritically, are most vulnerable and would benefit from automatically highlighting evidence in key responses and adding visual reliability indicators to reduce dependency on user prompts for verification.
\textit{Balanced Explorers}, more critical of outputs, reflected on responses, being more likely to spot hallucinations.
\textit{Targeted Explorers}, characterized by skepticism, requested evidence from the policy text as validation and are least affected.
Retrieval-Augmented Generation (RAG) could effectively reduce hallucinations.

\textbf{Further Considerations: }
Incomplete scraping may result in LLM responses based on partial policies, without alerting users -- adding ML-based techniques for policy classification~\cite{hosseini2021unifying} could enhance text integrity. Including HTML code and consent dialogues as LLM inputs could enable step-by-step privacy instructions.
Future work could explore open-source and locally run LLMs, as privacy interests themselves are sensitive.
Methodologically, Scenario 1 and Scenario 2 only provide exemplary settings for in-depth exploration or policy comparison. Both may vary in a different setting and may differ from the participants' use cases for the tool. Scenario 3 aims to capture participants' use cases for the tool, but the lab environment may not accurately reflect their daily life usage behavior. 
Our sample size limits generalizability. The four identified user profiles may not be exhaustive and need further exploration. A follow-up \textit{in-the-wild study} could provide deeper insights into real-world user behavior and tool performance.
In future work, we will also include awareness and comprehension tests for more objective results.

\section{Conclusion}
\label{sec:conclusion}

In the era of AI-driven hyper-personalization, increasing data collection heightens privacy risks, necessitating tools to make complex privacy policies accessible. Existing solutions often fail to improve privacy awareness and achieve understandable, efficient communication of privacy policies in a usable manner. 
We provide the first qualitative user study (N=22) on personalized LLM-driven privacy policy assessment. For this study, we have developed a prototype of such a tool as a Chrome extension. We find enhanced privacy awareness and understanding as perceived by users, 
but also concerns regarding LLM hallucinations and lacking policy evidence. 
Our findings offer guidance for future LLM-based privacy solutions.

\begin{acks}
The authors acknowledge the financial support by the Federal Ministry of Education and Research of Germany 
and by the Sächsische Staatsministerium für Wissenschaft Kultur und Tourismus in the program Center of Excellence for AI-research 
"Center for Scalable Data Analytics and Artificial Intelligence Dresden/Leipzig", 
project identification number: ScaDS.
\end{acks}

\bibliographystyle{ACM-Reference-Format}
\bibliography{literature}


\begin{thebibliography}{55}

%%% the Web address.

\ifx \showCODEN    \undefined \def \showCODEN     #1{\unskip}     \fi
\ifx \showDOI      \undefined \def \showDOI       #1{#1}\fi
\ifx \showISBNx    \undefined \def \showISBNx     #1{\unskip}     \fi
\ifx \showISBNxiii \undefined \def \showISBNxiii  #1{\unskip}     \fi
\ifx \showISSN     \undefined \def \showISSN      #1{\unskip}     \fi
\ifx \showLCCN     \undefined \def \showLCCN      #1{\unskip}     \fi
\ifx \shownote     \undefined \def \shownote      #1{#1}          \fi
\ifx \showarticletitle \undefined \def \showarticletitle #1{#1}   \fi
\ifx \showURL      \undefined \def \showURL       {\relax}        \fi
% invisible to TeX
\providecommand\bibfield[2]{#2}
\providecommand\bibinfo[2]{#2}
\providecommand\natexlab[1]{#1}
\providecommand\showeprint[2][]{arXiv:#2}

\bibitem[Amaral et~al\mbox{.}(2021)]%
        {amaral2021ai}
\bibfield{author}{\bibinfo{person}{Orlando Amaral}, \bibinfo{person}{Sallam Abualhaija}, \bibinfo{person}{Damiano Torre}, \bibinfo{person}{Mehrdad Sabetzadeh}, {and} \bibinfo{person}{Lionel~C Briand}.} \bibinfo{year}{2021}\natexlab{}.
\newblock \showarticletitle{AI-Enabled Automation for Completeness Checking of Privacy Policies}.
\newblock \bibinfo{journal}{\emph{IEEE Transactions on Software Engineering}} \bibinfo{volume}{48}, \bibinfo{number}{11} (\bibinfo{year}{2021}), \bibinfo{pages}{4647--4674}.
\newblock


\bibitem[Bartelt and Buchmann(2024)]%
        {transparency}
\bibfield{author}{\bibinfo{person}{Bianca Bartelt} {and} \bibinfo{person}{Erik Buchmann}.} \bibinfo{year}{2024}\natexlab{}.
\newblock \showarticletitle{Transparency in Privacy Policies}. In \bibinfo{booktitle}{\emph{12th International Conference on Building and Exploring Web Based Environments}}. \bibinfo{publisher}{IARIA Press}, \bibinfo{address}{Online}, \bibinfo{pages}{1--6}.
\newblock


\bibitem[Becher and Benoliel(2021)]%
        {becher2021law}
\bibfield{author}{\bibinfo{person}{Shmuel~I Becher} {and} \bibinfo{person}{Uri Benoliel}.} \bibinfo{year}{2021}\natexlab{}.
\newblock \showarticletitle{Law in Books and Law in Action: The Readability of Privacy Policies and the {GDPR}}. In \bibinfo{booktitle}{\emph{Consumer law and economics}}. \bibinfo{publisher}{Springer International Publishing}, \bibinfo{address}{Cham}, \bibinfo{pages}{179--204}.
\newblock


\bibitem[Belcheva et~al\mbox{.}(2023)]%
        {belcheva2023understanding}
\bibfield{author}{\bibinfo{person}{Veronika Belcheva}, \bibinfo{person}{Tatiana Ermakova}, {and} \bibinfo{person}{Benjamin Fabian}.} \bibinfo{year}{2023}\natexlab{}.
\newblock \showarticletitle{Understanding Website Privacy Policies—A Longitudinal Analysis Using Natural Language Processing}.
\newblock \bibinfo{journal}{\emph{Information}} \bibinfo{volume}{14}, \bibinfo{number}{11} (\bibinfo{year}{2023}), \bibinfo{pages}{622}.
\newblock


\bibitem[Benn and Lazar(2022)]%
        {benn2022s}
\bibfield{author}{\bibinfo{person}{Claire Benn} {and} \bibinfo{person}{Seth Lazar}.} \bibinfo{year}{2022}\natexlab{}.
\newblock \showarticletitle{What’s wrong with automated influence}.
\newblock \bibinfo{journal}{\emph{Canadian Journal of Philosophy}} \bibinfo{volume}{52}, \bibinfo{number}{1} (\bibinfo{year}{2022}), \bibinfo{pages}{125--148}.
\newblock


\bibitem[{California State Legislature}(2018)]%
        {CaliforniaStateLegislature2018}
\bibfield{author}{\bibinfo{person}{{California State Legislature}}.} \bibinfo{year}{2018}\natexlab{}.
\newblock \bibinfo{title}{AB-375 California Consumer Privacy Act of 2018}.
\newblock
\newblock
\urldef\tempurl%
\url{https://leginfo.legislature.ca.gov/faces/billTextClient.xhtml?bill_id=201720180AB375}
\showURL{%
\tempurl}
\newblock
\shownote{Accessed: Jul 2024}.


\bibitem[Chouaki et~al\mbox{.}(2022)]%
        {chouaki2022exploring}
\bibfield{author}{\bibinfo{person}{Salim Chouaki}, \bibinfo{person}{Islem Bouzenia}, \bibinfo{person}{Oana Goga}, {and} \bibinfo{person}{Beatrice Roussillon}.} \bibinfo{year}{2022}\natexlab{}.
\newblock \showarticletitle{Exploring the online micro-targeting practices of small, medium, and large businesses}.
\newblock \bibinfo{journal}{\emph{Proceedings of the ACM on Human-Computer Interaction}} \bibinfo{volume}{6}, \bibinfo{number}{CSCW2} (\bibinfo{year}{2022}), \bibinfo{pages}{1--23}.
\newblock


\bibitem[Contissa et~al\mbox{.}(2018)]%
        {contissa2018claudette}
\bibfield{author}{\bibinfo{person}{Giuseppe Contissa}, \bibinfo{person}{Koen Docter}, \bibinfo{person}{Francesca Lagioia}, \bibinfo{person}{Marco Lippi}, \bibinfo{person}{Hans-W Micklitz}, \bibinfo{person}{Przemys{\l}aw Pa{\l}ka}, \bibinfo{person}{Giovanni Sartor}, {and} \bibinfo{person}{Paolo Torroni}.} \bibinfo{year}{2018}\natexlab{}.
\newblock \bibinfo{title}{Claudette meets gdpr: Automating the evaluation of privacy policies using artificial intelligence}.
\newblock
\newblock


\bibitem[Corbin and Strauss(1990)]%
        {corbin1990grounded}
\bibfield{author}{\bibinfo{person}{Juliet Corbin} {and} \bibinfo{person}{Anselm Strauss}.} \bibinfo{year}{1990}\natexlab{}.
\newblock \showarticletitle{Grounded theory research: Procedures, canons, and evaluative criteria}.
\newblock \bibinfo{journal}{\emph{Qualitative sociology}} \bibinfo{volume}{13}, \bibinfo{number}{1} (\bibinfo{year}{1990}), \bibinfo{pages}{3--21}.
\newblock


\bibitem[Corbin and Strauss(2015)]%
        {corbin2015basics}
\bibfield{author}{\bibinfo{person}{Juliet Corbin} {and} \bibinfo{person}{Anselm Strauss}.} \bibinfo{year}{2015}\natexlab{}.
\newblock \bibinfo{booktitle}{\emph{Basics of qualitative research techniques and procedures for developing grounded theory}}.
\newblock \bibinfo{publisher}{Sage publications}, \bibinfo{address}{Thousand Oaks, CA, USA}.
\newblock


\bibitem[Cranor(2002)]%
        {cranor2002web}
\bibfield{author}{\bibinfo{person}{Lorrie Cranor}.} \bibinfo{year}{2002}\natexlab{}.
\newblock \bibinfo{booktitle}{\emph{Web privacy with P3P}}.
\newblock \bibinfo{publisher}{" O'Reilly Media, Inc."}, \bibinfo{address}{USA}.
\newblock


\bibitem[Cranor et~al\mbox{.}(2006)]%
        {cranor2006user}
\bibfield{author}{\bibinfo{person}{Lorrie~Faith Cranor}, \bibinfo{person}{Praveen Guduru}, {and} \bibinfo{person}{Manjula Arjula}.} \bibinfo{year}{2006}\natexlab{}.
\newblock \showarticletitle{User interfaces for privacy agents}.
\newblock \bibinfo{journal}{\emph{ACM Transactions on Computer-Human Interaction (TOCHI)}} \bibinfo{volume}{13}, \bibinfo{number}{2} (\bibinfo{year}{2006}), \bibinfo{pages}{135--178}.
\newblock


\bibitem[{European Union}(2016)]%
        {eu2016regulation}
\bibfield{author}{\bibinfo{person}{{European Union}}.} \bibinfo{year}{2016}\natexlab{}.
\newblock \showarticletitle{REGULATION {(EU) 2016/679 of the European Parliament} AND OF THE {Council} of 27 April 2016 on the protection of natural persons with regard to the processing of personal data and on the free movement of such data, and repealing {Directive 95/46/EC (General Data Protection Regulation)}}.
\newblock \bibinfo{journal}{\emph{Official Journal of the European Union}}  \bibinfo{volume}{L119/1} (\bibinfo{year}{2016}), \bibinfo{pages}{1--88}.
\newblock


\bibitem[Freiberger and Buchmann(2024)]%
        {freiberger2024legal}
\bibfield{author}{\bibinfo{person}{Vincent Freiberger} {and} \bibinfo{person}{Erik Buchmann}.} \bibinfo{year}{2024}\natexlab{}.
\newblock \showarticletitle{Legally Binding but Unfair? Towards Assessing Fairness of Privacy Policies}. In \bibinfo{booktitle}{\emph{Proceedings of the 10th ACM International Workshop on Security and Privacy Analytics}} \emph{(\bibinfo{series}{IWSPA '24})}. \bibinfo{publisher}{Association for Computing Machinery}, \bibinfo{address}{New York, NY, USA}, \bibinfo{pages}{15–22}.
\newblock


\bibitem[Gerber et~al\mbox{.}(2019a)]%
        {gerber2019investigating}
\bibfield{author}{\bibinfo{person}{Nina Gerber}, \bibinfo{person}{Benjamin Reinheimer}, {and} \bibinfo{person}{Melanie Volkamer}.} \bibinfo{year}{2019}\natexlab{a}.
\newblock \showarticletitle{Investigating people’s privacy risk perception}. In \bibinfo{booktitle}{\emph{Proceedings on Privacy Enhancing Technologies}}. \bibinfo{publisher}{PET Symposium}, \bibinfo{address}{Rochester, NY, USA}, \bibinfo{pages}{267--288}.
\newblock


\bibitem[Gerber et~al\mbox{.}(2019b)]%
        {gerber2019johnny}
\bibfield{author}{\bibinfo{person}{Nina Gerber}, \bibinfo{person}{Verena Zimmermann}, {and} \bibinfo{person}{Melanie Volkamer}.} \bibinfo{year}{2019}\natexlab{b}.
\newblock \showarticletitle{Why johnny fails to protect his privacy}. In \bibinfo{booktitle}{\emph{2019 IEEE European Symposium on Security and Privacy Workshops (EuroS\&PW)}}. \bibinfo{publisher}{IEEE}, \bibinfo{address}{Piscataway, NJ, USA}, \bibinfo{pages}{109--118}.
\newblock


\bibitem[Goram et~al\mbox{.}(2023)]%
        {goram2023human}
\bibfield{author}{\bibinfo{person}{Mandy Goram}, \bibinfo{person}{Tobias Dehling}, \bibinfo{person}{Felix Morsbach}, {and} \bibinfo{person}{Ali Sunyaev}.} \bibinfo{year}{2023}\natexlab{}.
\newblock \showarticletitle{Human-Centered Design for Data-Sparse Tailored Privacy Information Provision}.
\newblock In \bibinfo{booktitle}{\emph{Human Factors in Privacy Research}}. \bibinfo{publisher}{Springer International Publishing}, \bibinfo{address}{Cham}, \bibinfo{pages}{283--298}.
\newblock


\bibitem[Gr{\"u}newald et~al\mbox{.}(2023)]%
        {grunewald2023enabling}
\bibfield{author}{\bibinfo{person}{Elias Gr{\"u}newald}, \bibinfo{person}{Johannes~M Halkenh{\"a}u{\ss}er}, \bibinfo{person}{Nicola Leschke}, \bibinfo{person}{Johanna Washington}, \bibinfo{person}{Cristina Paupini}, {and} \bibinfo{person}{Frank Pallas}.} \bibinfo{year}{2023}\natexlab{}.
\newblock \showarticletitle{Enabling versatile privacy interfaces using machine-readable transparency information}. In \bibinfo{booktitle}{\emph{Privacy Symposium: Data Protection Law International Convergence and Compliance with Innovative Technologies}}. \bibinfo{publisher}{Springer International Publishing}, \bibinfo{address}{Cham}, \bibinfo{pages}{119--137}.
\newblock


\bibitem[Guo et~al\mbox{.}(2020)]%
        {guo2020poli}
\bibfield{author}{\bibinfo{person}{Wentao Guo}, \bibinfo{person}{Jay Rodolitz}, {and} \bibinfo{person}{Eleanor Birrell}.} \bibinfo{year}{2020}\natexlab{}.
\newblock \showarticletitle{Poli-see: An interactive tool for visualizing privacy policies}. In \bibinfo{booktitle}{\emph{Proceedings of the 19th Workshop on Privacy in the Electronic Society}}. \bibinfo{publisher}{Association for Computing Machinery}, \bibinfo{address}{New York, NY, USA}, \bibinfo{pages}{57--71}.
\newblock


\bibitem[Hamid et~al\mbox{.}(2023)]%
        {hamid2023genaipabench}
\bibfield{author}{\bibinfo{person}{Aamir Hamid}, \bibinfo{person}{Hemanth~Reddy Samidi}, \bibinfo{person}{Tim Finin}, \bibinfo{person}{Primal Pappachan}, {and} \bibinfo{person}{Roberto Yus}.} \bibinfo{year}{2023}\natexlab{}.
\newblock \bibinfo{title}{GenAIPABench: A benchmark for generative {AI-based} privacy assistants}.
\newblock \bibinfo{howpublished}{arXiv preprint arXiv:2309.05138}.
\newblock


\bibitem[Harkous et~al\mbox{.}(2018)]%
        {harkous2018polisis}
\bibfield{author}{\bibinfo{person}{Hamza Harkous}, \bibinfo{person}{Kassem Fawaz}, \bibinfo{person}{R{\'e}mi Lebret}, \bibinfo{person}{Florian Schaub}, \bibinfo{person}{Kang~G Shin}, {and} \bibinfo{person}{Karl Aberer}.} \bibinfo{year}{2018}\natexlab{}.
\newblock \showarticletitle{Polisis: Automated analysis and presentation of privacy policies using deep learning}. In \bibinfo{booktitle}{\emph{27th USENIX Security Symposium (USENIX Security 18)}}. \bibinfo{publisher}{USENIX Association}, \bibinfo{address}{USA}, \bibinfo{pages}{531--548}.
\newblock


\bibitem[Harkous et~al\mbox{.}(2016)]%
        {harkous2016pribots}
\bibfield{author}{\bibinfo{person}{Hamza Harkous}, \bibinfo{person}{Kassem Fawaz}, \bibinfo{person}{Kang~G Shin}, {and} \bibinfo{person}{Karl Aberer}.} \bibinfo{year}{2016}\natexlab{}.
\newblock \showarticletitle{{PriBots}: Conversational privacy with chatbots}. In \bibinfo{booktitle}{\emph{Twelfth Symposium on Usable Privacy and Security (SOUPS 2016)}}. \bibinfo{publisher}{USENIX Association}, \bibinfo{address}{USA}, \bibinfo{pages}{1--6}.
\newblock


\bibitem[Hosseini et~al\mbox{.}(2021)]%
        {hosseini2021unifying}
\bibfield{author}{\bibinfo{person}{Henry Hosseini}, \bibinfo{person}{Martin Degeling}, \bibinfo{person}{Christine Utz}, {and} \bibinfo{person}{Thomas Hupperich}.} \bibinfo{year}{2021}\natexlab{}.
\newblock \showarticletitle{Unifying privacy policy detection}.
\newblock \bibinfo{journal}{\emph{Proceedings on Privacy Enhancing Technologies}}  \bibinfo{volume}{4} (\bibinfo{year}{2021}), \bibinfo{pages}{480--499}.
\newblock


\bibitem[Jain et~al\mbox{.}(2021)]%
        {jain2021hyper}
\bibfield{author}{\bibinfo{person}{Geetika Jain}, \bibinfo{person}{Justin Paul}, {and} \bibinfo{person}{Archana Shrivastava}.} \bibinfo{year}{2021}\natexlab{}.
\newblock \showarticletitle{Hyper-personalization, co-creation, digital clienteling and transformation}.
\newblock \bibinfo{journal}{\emph{Journal of Business Research}}  \bibinfo{volume}{124} (\bibinfo{year}{2021}), \bibinfo{pages}{12--23}.
\newblock


\bibitem[Kelley et~al\mbox{.}(2009)]%
        {kelley2009nutrition}
\bibfield{author}{\bibinfo{person}{Patrick~Gage Kelley}, \bibinfo{person}{Joanna Bresee}, \bibinfo{person}{Lorrie~Faith Cranor}, {and} \bibinfo{person}{Robert~W Reeder}.} \bibinfo{year}{2009}\natexlab{}.
\newblock \showarticletitle{A" nutrition label" for privacy}. In \bibinfo{booktitle}{\emph{Proceedings of the 5th Symposium on Usable Privacy and Security}}. \bibinfo{publisher}{Association for Computing Machinery}, \bibinfo{address}{New York, NY, USA}, \bibinfo{pages}{1--12}.
\newblock


\bibitem[Kew et~al\mbox{.}(2023)]%
        {kew-etal-2023-bless}
\bibfield{author}{\bibinfo{person}{Tannon Kew}, \bibinfo{person}{Alison Chi}, \bibinfo{person}{Laura V{\'a}squez-Rodr{\'i}guez}, \bibinfo{person}{Sweta Agrawal}, \bibinfo{person}{Dennis Aumiller}, \bibinfo{person}{Fernando Alva-Manchego}, {and} \bibinfo{person}{Matthew Shardlow}.} \bibinfo{year}{2023}\natexlab{}.
\newblock \showarticletitle{{BLESS}: Benchmarking Large Language Models on Sentence Simplification}. In \bibinfo{booktitle}{\emph{Proceedings of the 2023 Conference on Empirical Methods in Natural Language Processing}}, \bibfield{editor}{\bibinfo{person}{Houda Bouamor}, \bibinfo{person}{Juan Pino}, {and} \bibinfo{person}{Kalika Bali}} (Eds.). \bibinfo{publisher}{Association for Computational Linguistics}, \bibinfo{address}{Singapore}, \bibinfo{pages}{13291--13309}.
\newblock


\bibitem[Kim et~al\mbox{.}(2024)]%
        {kim2024m}
\bibfield{author}{\bibinfo{person}{Sunnie~SY Kim}, \bibinfo{person}{Q~Vera Liao}, \bibinfo{person}{Mihaela Vorvoreanu}, \bibinfo{person}{Stephanie Ballard}, {and} \bibinfo{person}{Jennifer~Wortman Vaughan}.} \bibinfo{year}{2024}\natexlab{}.
\newblock \showarticletitle{" I'm Not Sure, But...": Examining the Impact of Large Language Models' Uncertainty Expression on User Reliance and Trust}. In \bibinfo{booktitle}{\emph{The 2024 ACM Conference on Fairness, Accountability, and Transparency}}. \bibinfo{publisher}{Association for Computing Machinery}, \bibinfo{address}{New York, NY, USA}, \bibinfo{pages}{822--835}.
\newblock


\bibitem[Kr{\"u}gel et~al\mbox{.}(2023)]%
        {krugel2023chatgpt}
\bibfield{author}{\bibinfo{person}{Sebastian Kr{\"u}gel}, \bibinfo{person}{Andreas Ostermaier}, {and} \bibinfo{person}{Matthias Uhl}.} \bibinfo{year}{2023}\natexlab{}.
\newblock \showarticletitle{{ChatGPT’s} inconsistent moral advice influences users’ judgment}.
\newblock \bibinfo{journal}{\emph{Scientific Reports}} \bibinfo{volume}{13}, \bibinfo{number}{1} (\bibinfo{year}{2023}), \bibinfo{pages}{4569}.
\newblock


\bibitem[Lee et~al\mbox{.}(2024)]%
        {lee2024deepfakes}
\bibfield{author}{\bibinfo{person}{Hao-Ping Lee}, \bibinfo{person}{Yu-Ju Yang}, \bibinfo{person}{Thomas~Serban Von~Davier}, \bibinfo{person}{Jodi Forlizzi}, {and} \bibinfo{person}{Sauvik Das}.} \bibinfo{year}{2024}\natexlab{}.
\newblock \showarticletitle{Deepfakes, Phrenology, Surveillance, and More! A Taxonomy of AI Privacy Risks}. In \bibinfo{booktitle}{\emph{Proceedings of the CHI Conference on Human Factors in Computing Systems}}. \bibinfo{publisher}{Association for Computing Machinery}, \bibinfo{address}{New York, NY, USA}, \bibinfo{pages}{1--19}.
\newblock


\bibitem[Malgieri(2020)]%
        {malgieri2020concept}
\bibfield{author}{\bibinfo{person}{Gianclaudio Malgieri}.} \bibinfo{year}{2020}\natexlab{}.
\newblock \showarticletitle{The Concept of Fairness in the {GDPR}: A Linguistic and Contextual Interpretation}. In \bibinfo{booktitle}{\emph{Proceedings of the 2020 Conference on fairness, accountability, and transparency}}. \bibinfo{publisher}{Association for Computing Machinery}, \bibinfo{address}{New York, NY, USA}, \bibinfo{pages}{154--166}.
\newblock


\bibitem[Martin(2022)]%
        {Martin2022}
\bibfield{author}{\bibinfo{person}{Kirsten Martin}.} \bibinfo{year}{2022}\natexlab{}.
\newblock \showarticletitle{Manipulation, Privacy, and Choice}.
\newblock \bibinfo{journal}{\emph{North Carolina Journal of Law \& Technology}}  \bibinfo{volume}{23} (\bibinfo{year}{2022}), \bibinfo{pages}{452}.
\newblock


\bibitem[Mhaidli et~al\mbox{.}(2023)]%
        {mhaidli2023researchers}
\bibfield{author}{\bibinfo{person}{Abraham Mhaidli}, \bibinfo{person}{Selin Fidan}, \bibinfo{person}{An Doan}, \bibinfo{person}{Gina Herakovic}, \bibinfo{person}{Mukund Srinath}, \bibinfo{person}{Lee Matheson}, \bibinfo{person}{Shomir Wilson}, {and} \bibinfo{person}{Florian Schaub}.} \bibinfo{year}{2023}\natexlab{}.
\newblock \showarticletitle{Researchers’ experiences in analyzing privacy policies: Challenges and opportunities}.
\newblock \bibinfo{journal}{\emph{Proceedings on Privacy Enhancing Technologies}}  \bibinfo{volume}{4} (\bibinfo{year}{2023}), \bibinfo{pages}{287--305}.
\newblock


\bibitem[Myers(2020)]%
        {myers2020qualitative}
\bibfield{author}{\bibinfo{person}{Michael~D Myers}.} \bibinfo{year}{2020}\natexlab{}.
\newblock \bibinfo{booktitle}{\emph{Qualitative research in business \& management}}.
\newblock \bibinfo{publisher}{Sage Publications Limited}, \bibinfo{address}{London, UK}.
\newblock


\bibitem[Nokhbeh~Zaeem et~al\mbox{.}(2020)]%
        {nokhbeh2020privacycheck}
\bibfield{author}{\bibinfo{person}{Razieh Nokhbeh~Zaeem}, \bibinfo{person}{Safa Anya}, \bibinfo{person}{Alex Issa}, \bibinfo{person}{Jake Nimergood}, \bibinfo{person}{Isabelle Rogers}, \bibinfo{person}{Vinay Shah}, \bibinfo{person}{Ayush Srivastava}, {and} \bibinfo{person}{K~Suzanne Barber}.} \bibinfo{year}{2020}\natexlab{}.
\newblock \showarticletitle{PrivacyCheck v2: A Tool that Recaps Privacy Policies for You}. In \bibinfo{booktitle}{\emph{Proceedings of the 29th ACM international conference on information \& knowledge management}}. \bibinfo{publisher}{Association for Computing Machinery}, \bibinfo{address}{New York, NY, USA}, \bibinfo{pages}{3441--3444}.
\newblock


\bibitem[Obar and Oeldorf-Hirsch(2020)]%
        {obar2020biggest}
\bibfield{author}{\bibinfo{person}{Jonathan~A Obar} {and} \bibinfo{person}{Anne Oeldorf-Hirsch}.} \bibinfo{year}{2020}\natexlab{}.
\newblock \showarticletitle{The biggest lie on the internet: Ignoring the privacy policies and terms of service policies of social networking services}.
\newblock \bibinfo{journal}{\emph{Information, Communication \& Society}} \bibinfo{volume}{23}, \bibinfo{number}{1} (\bibinfo{year}{2020}), \bibinfo{pages}{128--147}.
\newblock


\bibitem[OpenAI(2024a)]%
        {OpenAI2024}
\bibfield{author}{\bibinfo{person}{OpenAI}.} \bibinfo{year}{2024}\natexlab{a}.
\newblock \bibinfo{title}{GPT-4o}.
\newblock
\newblock
\urldef\tempurl%
\url{https://openai.com/index/hello-gpt-4o/}
\showURL{%
\tempurl}
\newblock
\shownote{Accessed: Jul 2024}.


\bibitem[OpenAI(2024b)]%
        {o_mini}
\bibfield{author}{\bibinfo{person}{OpenAI}.} \bibinfo{year}{2024}\natexlab{b}.
\newblock \bibinfo{title}{GPT-4o mini: advancing cost-efficient intelligence}.
\newblock \bibinfo{howpublished}{\url{https://openai.com/index/gpt-4o-mini-advancing-cost-efficient-intelligence/}}.
\newblock
\newblock
\shownote{Accessed: 2025-01-13}.


\bibitem[{OpenAI}(2024)]%
        {openai2024whisperlargev3}
\bibfield{author}{\bibinfo{person}{{OpenAI}}.} \bibinfo{year}{2024}\natexlab{}.
\newblock \bibinfo{title}{{whisper-large-v3}}.
\newblock \bibinfo{howpublished}{\url{https://huggingface.co/openai/whisper-large-v3}}.
\newblock
\newblock
\shownote{Accessed: 2024-09-02}.


\bibitem[Pollach(2005)]%
        {pollach2005typology}
\bibfield{author}{\bibinfo{person}{Irene Pollach}.} \bibinfo{year}{2005}\natexlab{}.
\newblock \showarticletitle{A Typology of Communicative Strategies in Online Privacy Policies: Ethics, Power and Informed Consent}.
\newblock \bibinfo{journal}{\emph{Journal of Business Ethics}}  \bibinfo{volume}{62} (\bibinfo{year}{2005}), \bibinfo{pages}{221--235}.
\newblock


\bibitem[Ravichander et~al\mbox{.}(2021)]%
        {Ravichander2021}
\bibfield{author}{\bibinfo{person}{Abhilasha Ravichander}, \bibinfo{person}{Alan~W. Black}, \bibinfo{person}{Thomas Norton}, \bibinfo{person}{Shomir Wilson}, {and} \bibinfo{person}{Norman Sadeh}.} \bibinfo{year}{2021}\natexlab{}.
\newblock \showarticletitle{Breaking Down Walls of Text: How Can NLP Benefit Consumer Privacy?}. In \bibinfo{booktitle}{\emph{ACL '21: Annual Meeting of the Association for Computational Linguistics}}. \bibinfo{publisher}{Association for Computational Linguistics}, \bibinfo{address}{Online}, \bibinfo{pages}{4125--4140}.
\newblock


\bibitem[Reidenberg et~al\mbox{.}(2015)]%
        {reidenberg2015disagreeable}
\bibfield{author}{\bibinfo{person}{Joel~R Reidenberg}, \bibinfo{person}{Travis Breaux}, \bibinfo{person}{Lorrie~Faith Cranor}, \bibinfo{person}{Brian French}, \bibinfo{person}{Amanda Grannis}, \bibinfo{person}{James~T Graves}, \bibinfo{person}{Fei Liu}, \bibinfo{person}{Aleecia McDonald}, \bibinfo{person}{Thomas~B Norton}, \bibinfo{person}{Rohan Ramanath}, {et~al\mbox{.}}} \bibinfo{year}{2015}\natexlab{}.
\newblock \showarticletitle{Disagreeable privacy policies: Mismatches between meaning and users' understanding}.
\newblock \bibinfo{journal}{\emph{Berkeley Tech. LJ}}  \bibinfo{volume}{30} (\bibinfo{year}{2015}), \bibinfo{pages}{39}.
\newblock


\bibitem[{remram44}(2024)]%
        {remram44_taguette}
\bibfield{author}{\bibinfo{person}{{remram44}}.} \bibinfo{year}{2024}\natexlab{}.
\newblock \bibinfo{title}{{Taguette}}.
\newblock \bibinfo{howpublished}{\url{https://gitlab.com/remram44/taguette}}.
\newblock
\newblock
\shownote{Accessed: 2024-09-02}.


\bibitem[Rodriguez et~al\mbox{.}(2024)]%
        {rodriguez2024large}
\bibfield{author}{\bibinfo{person}{David Rodriguez}, \bibinfo{person}{Ian Yang}, \bibinfo{person}{Jose~M Del~Alamo}, {and} \bibinfo{person}{Norman Sadeh}.} \bibinfo{year}{2024}\natexlab{}.
\newblock \showarticletitle{Large language models: a new approach for privacy policy analysis at scale}.
\newblock \bibinfo{journal}{\emph{Computing}}  \bibinfo{volume}{106} (\bibinfo{year}{2024}), \bibinfo{pages}{1--25}.
\newblock


\bibitem[Roy(2024)]%
        {Roy2024}
\bibfield{author}{\bibinfo{person}{Hugo Roy}.} \bibinfo{year}{2024}\natexlab{}.
\newblock \bibinfo{title}{Terms of Service; Didn’t Read}.
\newblock
\newblock
\urldef\tempurl%
\url{https://tosdr.org/}
\showURL{%
\tempurl}
\newblock
\shownote{Accessed: 2024-09-03}.


\bibitem[Schaub et~al\mbox{.}(2017)]%
        {schaub2017designing}
\bibfield{author}{\bibinfo{person}{Florian Schaub}, \bibinfo{person}{Rebecca Balebako}, {and} \bibinfo{person}{Lorrie~Faith Cranor}.} \bibinfo{year}{2017}\natexlab{}.
\newblock \showarticletitle{Designing effective privacy notices and controls}.
\newblock \bibinfo{journal}{\emph{IEEE Internet Computing}} \bibinfo{volume}{21}, \bibinfo{number}{3} (\bibinfo{year}{2017}), \bibinfo{pages}{70--77}.
\newblock


\bibitem[Schaub et~al\mbox{.}(2016)]%
        {schaub2016watching}
\bibfield{author}{\bibinfo{person}{Florian Schaub}, \bibinfo{person}{Aditya Marella}, \bibinfo{person}{Pranshu Kalvani}, \bibinfo{person}{Blase Ur}, \bibinfo{person}{Chao Pan}, \bibinfo{person}{Emily Forney}, {and} \bibinfo{person}{Lorrie~Faith Cranor}.} \bibinfo{year}{2016}\natexlab{}.
\newblock \showarticletitle{Watching them watching me: Browser extensions’ impact on User Privacy Awareness and concern}. In \bibinfo{booktitle}{\emph{NDSS workshop on usable security}}, Vol.~\bibinfo{volume}{10}. \bibinfo{publisher}{The Internet Society}, \bibinfo{address}{Reston, VA, USA}, \bibinfo{pages}{1--10}.
\newblock


\bibitem[Steinfeld(2016)]%
        {steinfeld2016agree}
\bibfield{author}{\bibinfo{person}{Nili Steinfeld}.} \bibinfo{year}{2016}\natexlab{}.
\newblock \showarticletitle{“I agree to the terms and conditions”:(How) do users read privacy policies online? An eye-tracking experiment}.
\newblock \bibinfo{journal}{\emph{Computers in human behavior}}  \bibinfo{volume}{55} (\bibinfo{year}{2016}), \bibinfo{pages}{992--1000}.
\newblock


\bibitem[SYSTRAN(2023)]%
        {systran2023fasterwhisper}
\bibfield{author}{\bibinfo{person}{SYSTRAN}.} \bibinfo{year}{2023}\natexlab{}.
\newblock \bibinfo{title}{{faster-whisper}}.
\newblock \bibinfo{howpublished}{\url{https://github.com/SYSTRAN/faster-whisper}}.
\newblock
\newblock
\shownote{Accessed: 2024-09-02}.


\bibitem[Tankelevitch et~al\mbox{.}(2024)]%
        {tankelevitch2024metacognitive}
\bibfield{author}{\bibinfo{person}{Lev Tankelevitch}, \bibinfo{person}{Viktor Kewenig}, \bibinfo{person}{Auste Simkute}, \bibinfo{person}{Ava~Elizabeth Scott}, \bibinfo{person}{Advait Sarkar}, \bibinfo{person}{Abigail Sellen}, {and} \bibinfo{person}{Sean Rintel}.} \bibinfo{year}{2024}\natexlab{}.
\newblock \showarticletitle{The metacognitive demands and opportunities of generative AI}. In \bibinfo{booktitle}{\emph{Proceedings of the CHI Conference on Human Factors in Computing Systems}}. \bibinfo{publisher}{Association for Computing Machinery}, \bibinfo{address}{New York, NY, USA}, \bibinfo{pages}{1--24}.
\newblock


\bibitem[Tesfay et~al\mbox{.}(2018)]%
        {tesfay2018privacyguide}
\bibfield{author}{\bibinfo{person}{Welderufael~B Tesfay}, \bibinfo{person}{Peter Hofmann}, \bibinfo{person}{Toru Nakamura}, \bibinfo{person}{Shinsaku Kiyomoto}, {and} \bibinfo{person}{Jetzabel Serna}.} \bibinfo{year}{2018}\natexlab{}.
\newblock \showarticletitle{PrivacyGuide: Towards an Implementation of the EU GDPR on Internet Privacy Policy Evaluation}. In \bibinfo{booktitle}{\emph{Proceedings of the Fourth ACM International Workshop on Security and Privacy Analytics}}. \bibinfo{publisher}{Association for Computing Machinery}, \bibinfo{address}{New York, NY, USA}, \bibinfo{pages}{15--21}.
\newblock


\bibitem[Torre et~al\mbox{.}(2020)]%
        {torre2020ai}
\bibfield{author}{\bibinfo{person}{Damiano Torre}, \bibinfo{person}{Sallam Abualhaija}, \bibinfo{person}{Mehrdad Sabetzadeh}, \bibinfo{person}{Lionel Briand}, \bibinfo{person}{Katrien Baetens}, \bibinfo{person}{Peter Goes}, {and} \bibinfo{person}{Sylvie Forastier}.} \bibinfo{year}{2020}\natexlab{}.
\newblock \showarticletitle{An ai-assisted approach for checking the completeness of privacy policies against gdpr}. In \bibinfo{booktitle}{\emph{2020 IEEE 28th International Requirements Engineering Conference (RE)}}. \bibinfo{publisher}{IEEE}, \bibinfo{address}{Piscataway, NJ, USA}, \bibinfo{pages}{136--146}.
\newblock


\bibitem[Windl et~al\mbox{.}(2022)]%
        {windl2022automating}
\bibfield{author}{\bibinfo{person}{Maximiliane Windl}, \bibinfo{person}{Niels Henze}, \bibinfo{person}{Albrecht Schmidt}, {and} \bibinfo{person}{Sebastian~S Feger}.} \bibinfo{year}{2022}\natexlab{}.
\newblock \showarticletitle{Automating contextual privacy policies: Design and evaluation of a production tool for digital consumer privacy awareness}. In \bibinfo{booktitle}{\emph{Proceedings of the 2022 CHI Conference on Human Factors in Computing Systems}}. \bibinfo{publisher}{Association for Computing Machinery}, \bibinfo{address}{New York, NY, USA}, \bibinfo{pages}{1--18}.
\newblock


\bibitem[Woodring et~al\mbox{.}(2024)]%
        {woodring2024enhancing}
\bibfield{author}{\bibinfo{person}{Justin Woodring}, \bibinfo{person}{Katherine Perez}, {and} \bibinfo{person}{Aisha Ali-Gombe}.} \bibinfo{year}{2024}\natexlab{}.
\newblock \showarticletitle{Enhancing privacy policy comprehension through Privacify: A user-centric approach using advanced language models}.
\newblock \bibinfo{journal}{\emph{Computers \& Security}}  \bibinfo{volume}{145} (\bibinfo{year}{2024}), \bibinfo{pages}{103997}.
\newblock


\bibitem[Xiang et~al\mbox{.}(2023)]%
        {xiang2023policychecker}
\bibfield{author}{\bibinfo{person}{Anhao Xiang}, \bibinfo{person}{Weiping Pei}, {and} \bibinfo{person}{Chuan Yue}.} \bibinfo{year}{2023}\natexlab{}.
\newblock \showarticletitle{PolicyChecker: Analyzing the GDPR Completeness of Mobile Apps' Privacy Policies}. In \bibinfo{booktitle}{\emph{Proceedings of the 2023 ACM SIGSAC Conference on Computer and Communications Security}}. \bibinfo{publisher}{Association for Computing Machinery}, \bibinfo{address}{New York, NY, USA}, \bibinfo{pages}{3373--3387}.
\newblock


\bibitem[Zaeem and Barber(2020)]%
        {Zaeem2020}
\bibfield{author}{\bibinfo{person}{Razieh~Nokhbeh Zaeem} {and} \bibinfo{person}{K.~Suzanne Barber}.} \bibinfo{year}{2020}\natexlab{}.
\newblock \showarticletitle{The Effect of the {GDPR} on Privacy Policies: Recent Progress and Future Promise}.
\newblock \bibinfo{journal}{\emph{ACM Trans. Manage. Inf. Syst.}} \bibinfo{volume}{12}, \bibinfo{number}{1}, Article \bibinfo{articleno}{2} (\bibinfo{date}{dec} \bibinfo{year}{2020}), \bibinfo{numpages}{20}~pages.
\newblock


\end{thebibliography}
\appendix
\section{Appendix}
\label{sec:appendix}
\subsection{Prompting}\label{app:prompting}
\subsubsection{Prompting Approach for Initial Assessment Generation}

    Your output must be a maximum of 600 words long! You are an expert in data protection and a member of an ethics council. You are given a privacy policy. Your task is to uncover aspects in data protection declarations that are ethically questionable from your perspective. Proceed \textbf{step by step}:

    \begin{enumerate}
        \item \textbf{Criteria:} From your perspective, identify relevant ethical test criteria for this privacy policy as criteria for a later evaluation. When naming the test criteria, stick to standardized terms and concepts that are common in the field of ethics. Keep it short! 
        \item \textbf{Analysis:} Based on this, check for ethical problems or ethically questionable circumstances in the privacy policy. 
        \item \textbf{Evaluation:} Only after you have completed step 2: Rate the privacy policy based on your analysis regarding each of your criteria on a 5-point Likert scale. Explain what this rating means. Explain what the ideal case with 5 points and the worst case with one point would look like. The output in this step should look like this: 
        [Insert rating criterion here]: [insert rating here]/5 [insert line break] 
        [insert justification here]
        
        \item \textbf{Conclusion:} Reflect on your evaluation and check whether it is complete.
    \end{enumerate}
    
    Important: Check for errors in your analysis and correct them if necessary before the evaluation. You must present your approach clearly and concisely and follow the steps mentioned. Your output must not exceed 600 words.

\subsubsection{Prompting Approach for Chat Answer Generation}
\label{sec:chat_prompt}

\textbf{System prompt criteria chat: }Keep it short! Privacy policy: <Privacy policy here> | Rating: <criteria evaluation result here>. Users want to know more about how this rating is justified in the privacy policy. When answering the questions, focus on the given topic of the rating. Keep it short! <Complexity and answer length according to settings>
\\
\textbf{System prompt general chat: }You are an expert in data protection with many years of experience in consumer protection. You have analyzed the following privacy policy and are aware of its risks and ethical implications for users: <Privacy policy here>. 
You should advise users and explain the implications for them in a conversation. <Complexity and answer length according to settings>

\subsubsection{Prompting Approach for our Suggested Question Generation}
\label{sec:suggestion_prompt}

\textbf{System prompt: }Your task is to ask questions about a privacy policy. Your output consists of three questions: 1. question 1; 2. question 2; 3. question 3. Please output the questions in a numbered list. Never repeat questions that have already been asked: <already asked questions here>
\textbf{User prompt: }Specifically: Ask your questions about the privacy policy on the topic: <criterion inserted here>.
Embrace the context of the previous chat: <chat history here>

   \subsection{Scenario Description}
\label{sec:scenario}
    We provide you with a browser extension, which you are to use to find out about the privacy practices of websites.
Over the next 20 minutes, you will work through these scenarios on your own. The 20 minutes refers to all scenarios together. If you need help, have a question for us or are stuck, please let us know.

\textbf{Scenario 1:}
You would like to find out about current world events. You regularly read Focus and would like to try out their digital offering. You find out how [news portal in the authors' country] handles data protection: 
\url{https://www.focus.de/} 
Take the necessary time to go into detail. 
What is your most important realisation?
\\
Starting from the fact that you now want to make a digital subscription:
Now you need to choose a payment method. You are thinking about setting up an account with PayPal. You use the tool to find out about PayPal's data protection practices. 
\url{https://www.paypal.com/de/digital-wallet/ways-to-pay/checkout-with-paypal}  
Take the necessary time to go into detail.
What is your most important realisation?

\textbf{Scenario 2:}
Imagine you are in the process of shopping online. You want to buy books. In your search on the Internet, you come across various web shops. Assume that at least the address, contact details and payment information are required for a purchase: 
\url{https://www.hugendubel.de/de}, \url{https://www.buchkatalog.de/}, \url{https://www.amazon.de/ref=nav_logo?language=de_DE&currency=EUR}, 
\\
and \url{https://www.kopp-verlag.de/}. 
For each of these sites, you must consider whether you agree with a purchase and thus the data protection standard of the sites. Use the application to explore the sites with regard to your personal data protection preferences. You are also welcome to compare the sites: 
Ratings [Shop 1-4]: \\
Where would you be most likely to buy? Why? 

\textbf{Scenario 3: }
Free exploration: search freely for websites where you want to find out about data protection practices. Use the remaining time to explore freely as if you were at home. Let your curiosity run free.

\clearpage

\subsection{Interview Guide}
\label{sec:guide}

\begin{table}[!ht]
\small

\begin{tabular}{|p{12.7cm}|p{4.0cm}|}

\hline
\textbf{Exemplary Questions}  & \textbf{Expected} \textbf{Results}\\
\hline

What were your first impressions of the extension?\par 
Was there anything you found surprising or unexpected?\par 
Did you face any issues using it?\par 
How did the information presented by the extension make you feel? &  First thoughts and impressions, deviations from expectations  \\
\hline

Did you miss any information being presented in the application?\par 
  How clear was the language used in the application? Were there any terms, phrases, or instructions that you found confusing or unclear?\par Were there moments when you felt overwhelmed by the information presented? \par 
  In case you have specific accessibility needs (e.g. vision or hearing impairment): How well did the extension accommodate this? & Evaluation of \ref{item:rq-understandable} \\
\hline

Do you feel like having a better understanding of the issues regarding data protection such a website can have? Explain!\par Did the extension make you aware of any privacy-related issues you were not previously aware of? If yes, can you describe these issues?\par Would you consider changing any of your browsing habits using this extension? If so, how? & Evaluation of \ref{item:rq-awareness} \\
\hline

How quickly were you able to find the information of interest using the extension?\par  How effective was the extension in helping you get an overview regarding data protection on the websites?\par
What are your thoughts on the overall design of the extension's interface?\par 
 Were there any features or design elements that you found unnecessary or confusing?\par 
 Is there any information the extension missed or did not highlight enough or not in the right presentation style? \par
What changes would you make? \par 
What aspects of the extension contributed most to your satisfaction or dissatisfaction? & Evaluation of \ref{item:rq-usability}\\
\hline

Is there anything else you’d like to share about your experience using the extension?  & Open issues  \\
\hline

\end{tabular}
\label{tab:guide}

\end{table}

\subsection{Issues in LLM Chat Responses}
\label{sec:chat_res}
\begin{table}[!ht]
\begin{tabular}{|p{2.5cm}|p{1.9cm}|p{12cm}|}
\hline
\textbf{Problem}&\textbf{Occurrences} &\textbf{Example}\\
\hline

Incompletely Scraped Policies & 13 (4 policies) & For an incompletely scraped policy: \enquote{[...]There are no clear details about the security measures.}\\
\hline

Context-related Limitations &	4 & \textbf{Question:} \enquote{What is the link for this form?}\par\textbf{Response:} \enquote{The specific link for the [...] online data protection request form is not provided directly in the privacy policy. [...]}\\
\hline

Overly Euphemistic Statements &	12 & \textbf{Question:} \enquote{Where will my data be transmitted to?} \par\textbf{Response:} \enquote{[...] To summarize: Your data is mainly shared within TikTok and with trusted partners to operate and improve the app.}\\
\hline

Omission of Information &	12 & \textbf{Question:} \enquote{[...] and how is this communicated in the privacy policy?} \par\textbf{Response:} \enquote{[...] This is explained in the privacy policy under the relevant sections.}\\
\hline

Generic Responses &	8 & \textbf{Question:} \enquote{What options do users have to withdraw their consent at any time and how is this process communicated?} \par\textbf{Response:} \enquote{Users can withdraw their consent at any time by changing certain settings in their account or by contacting customer service. This process is explained in the privacy policy. [...]}\\
\hline

Hallucinations &	5 & \textbf{Question:} \enquote{What's the worst privacy policy you've read so far?}\par
\textbf{Response:} \enquote{[...][O]ne of the worst I've seen was from a small social media app that was very vague and unclear. [...]}\\
\hline

Misunderstanding User Queries &	4 & \textbf{Question:} \enquote{Please show me the entirety of the categories to evaluate privacy policies.}\par \textbf{Response:} \enquote{Of course, I will summarize the most important categories and aspects of the privacy policy [...].}\\
\hline

Mixed Signals &	2 & \textbf{Question:} \enquote{Is only the most necessary data collected?} \par\textbf{Response:} \enquote{No, PayPal collects a variety of data categories, including sensitive data. Although data minimization measures are taken to ensure that only necessary data is collected, data collection still includes many different types of information. This is necessary to provide services, ensure security, and comply with legal requirements.}\\
\hline

\end{tabular}  %
\label{tab:answer_issues}
\end{table}

\clearpage

\subsection{Codebook}
\label{sec:codebook}
\begin{table}[!ht]
\small
\begin{tabular}{|p{3.5 cm}|p{11.5 cm}|p{1.3 cm}|}
\hline
\textbf{Code}                                                   & \textbf{Description}                                                                                                                                                                     & \textbf{\#Passages} \\ \hline
\multicolumn{3}{|p{14.3 cm}|}{\textbf{Topic 1:} User Attitudes, Motivations, and Behavior
}\\
\hline

indifference                                          & Data protection issues do not concern participants or are of no interest to them                                                                                              & 7                    
\\\hline

emotional response                                    & Tool triggers emotional response of participants                                                                                                                                & 7                    \\ \hline

professional use cases                                & Usecases of the tool that facilitate business processes                                                                                                                         & 10                   \\ \hline

existing privacy   misconceptions                     & Participants having privacy misconceptions or distorted view of reality   regarding privacy                                                                                     & 4                    \\ \hline

multifactorial \&   context-dependent decision making & Participants note that their decisions are based on multiple factors   aside from data protection and depend on the given context                                               & 11\\
\hline

habitualized behavior                                 & Participants are unwilling to change behavior due to existing   habitualized behavior and inconvenience of a change                                                             & 15                   \\ 
\hline

curiosity-driven use                                  & Participants explore using the tool out of curiosity                                                                                                                            & 9                    \\ 
\hline

personal usage interest                               & Participants want to use the tool in their daily personal lives                                                                                                                 & 15                   \\ \hline

insecurity regarding data   protection                & Participants feel insecure about data protection issues                                                                                                                         & 3                    \\ \hline

behavior depends on setting of   use                  & Depending on the given setting like public or own computer use the usage   pattern may differ                                                                                  & 1  \\
\hline

negative predisposition and   distrust                & Participants expect a very low standard of data protection from websites,   are negatively biased by a website's design or have general distrust in   websites' data protection & 30                   \\ \hline

\multicolumn{3}{|p{14.3 cm}|}{\textbf{Topic 2:} Information Quality and Clarity
}\\
\hline

language clarity and   simplicity                     & The used language is easy to understand, clear and simple                                                                                                                       & 35                   \\ \hline

quick and effective overview                          & The tool provides participants with a quick and effective overview of all   relevant information                                                                                & 51                   \\ \hline

Evaluation transparency                               & Aspects regarding how transparent the evaluation process is to users                                                                                                            & 22                   \\ \hline

Levels of detail                                      & Degree of detail and context provided and to what degree it   differentiates evaluations on the different levels of depth in the   application                                  & 35                   \\ \hline

answer quality                                        & Chat answers are helping participants effectively                                                                                                                               & 35                   \\ \hline

Communicated information is   incomplete              & The tool communicates to the user that there is no specific or vague   information on the topic in the policy                                                                   & 7                    \\ \hline

chat flexibility                                      & The chat handles typos, other languages, areas out of context or other   challenges                                                                                             & 6                    \\ \hline

chat consistency                                      & Conversations are consistent and continous in chat, between similar   policies the chat answers for the same questions are also similar                                         & 5                    \\ \hline

less vague \& more to the   point                     & Information presented by the tool should be less vague and more to the   point                                                                                                  & 20                   \\ \hline

adaptability with settings                            & Praise for adjustability of the tool by changing settings                                                                                                                       & 11                   \\ \hline

information rich                                      & The information presented by the application is plentiful, rich and   covering everything relevant                                                                              & 17                   \\ \hline

\multicolumn{3}{|p{14.3 cm}|}{\textbf{Topic 3:} User Experience and Interface Interaction}\\
\hline

good visual cue                                       & The initial smiley icon as visual cue is praised for its design,   placement and increased awareness                                                                            & 29                   \\ \hline

formatting and layout issues                          & Issues addressing the formatting and layout of the application eg. of the   chat output                                                                                         & 36                   \\ \hline

suggestion quality                                    & The quality of the chat query suggestions provided by the tool is praised for its   inspiring and guiding effect but criticised for being too long, unprecise and   not diverse enough     & 23                   \\ \hline

playful                                               & Participants perceive the tool as playful and fun                                                                                                                               & 2                    \\ \hline

visual cue issues                                     & Issues with the visual cue being to intrusive, emotionally loaded,   technichal issues with it due to not appearing, being covered, changing   colors and similar issues        & 22                   \\ \hline

easy and intuitive use                                & How easy and intuitive the application is to use                                                                                                                                & 51                   \\ 
\hline

Loading times                                         & Aspects addressing feedback when the application is loading and long   loading times                                                                                            & 11                   \\ 

\hline

More differentiated initial   scoring                 & More nuanced initial assessment scoring                                                                                                                                         & 4                    \\ 
\hline

Button usability                                      & Whether buttons are easy to use by giving enough feedback when clicked   and being named appropriately                                                                          & 14                   \\ \hline

navigation difficulties                               & Participants face difficulties in navigating the application mostly due   to scrolling                                                                                          & 18                   \\ \hline

challenging to ask precise   questions                & Participants struggle to form precise questions when they want to ask for   specific information                                                                                & 6                    \\ \hline

criteria dashboard landing   page                     & Put the overview dashboard on the default view                                                                                                                                  & 4                    \\ \hline

difficult-to-find or confusing   UI elements          & Including mix up of chat output and text field, terminology explanations   and policy text difficult to find                                                                    & 27                   \\ \hline

responsive                                            & The tool responds quickly                                                                                                                                                       & 8                    \\ \hline

accessibility features                                & voice input and audio output                                                                                                                                                    & 7                    \\ \hline

\end{tabular}
\end{table}
\clearpage
\begin{table}[h!]
\small
\begin{tabular}{|p{3.5 cm}|p{11.5 cm}|p{1.3 cm}|}
\hline
\textbf{Code}                                                   & \textbf{Description}                                                                                                                                                                     & \textbf{\#Passages} \\
                    \hline

\multicolumn{3}{|p{14.3 cm}|}{\textbf{Topic 4:} Tool Reliability and Trustworthiness}\\
\hline

hallucination risks and LLM   limitations             & Limitations in the LLM's factual accuracy                                                                                                                                       & 27                   \\ \hline

chat relativizes initial   assessment                 & Cases in which chat answers contradict the tool's initial assessment  to some degree                                                                                            & 6                    \\ \hline

trust issue in tool                                   & Participants express issues with trusting the results of the tool                                                                                                               & 17                   \\ \hline

scraper limitations                                   & The scraper used by the tool cannot access some pages privacy policies                                                                                                          & 9                    \\ \hline

inconsistent evaluation   criteria                    & Evaluation criteria are not fixed and change between policy assessments,   issues with comparibility entail                                                                     & 22                   \\ \hline

validate correct policy                               & The tool should validate whether the correct and full policy has been   scraped                                                                                                 & 13                   \\ \hline

rating accuracy                                       & The rating of a policy by the tool is accurate                                                                                                                                  & 4                    \\ \hline

account for differing   relevance of criteria& Not all evaluation criteria are equally relevant, which should be   considered in the assessment                                                                                & 5                    \\ \hline

policy evidence                                       & The tool should utilize quotes or links to the policy as evidence for   presented information                                                                                   & 14                   \\ \hline

\multicolumn{3}{|p{14.3 cm}|}{\textbf{Topic 5:} Recommendations for Features and Functional Improvements}\\
\hline

More customization options                            & Further settings allowing for a more customized use of the tool                                                                                                                 & 4                    \\ \hline

actionable solution                                   & Tool should provide actionable solutions like recommendations,   alternatives, automated adjustment of cookie settings,...                                                      & 13                   \\ \hline

side-by-side comparison                               & Participants wish for side by side comparisons between multiple pages                                                                                                           & 13                   \\ 
 \hline

broader focus on security   threats \& leaks          & The tool should cover cybersecurity more broadly and highlight recent   breaches                                                                                                & 4                    \\ \hline

communicate policy length /   complexity              & Provide background information on policy length and complexity                                                                                                                  & 1                    \\ \hline

contextual feedback privacy   setings                 & Provide feedback on current privacy settings and their context on the   given page                                                                                              & 4                    \\ \hline

read-only variant                                     & Proposed change of the tool to be read-only without interactive chat                                                                                                            & 1                    \\ \hline

multiple services involved                            & Typically not just one service is involved in facilitating the users'   goals and all involved services would need to be checked along the user   journey                       & 1                    \\ \hline

dark mode                                             & Dark mode                                                                                                                                                                       & 1                    \\ \hline

window size \& scaling                                & The window size and scaling of elements in the application should be   bigger and adaptable                                                                                     & 20                   \\ 
\hline

\multicolumn{3}{|p{14.3 cm}|}{\textbf{Topic 6:} Impact on Users}\\
\hline

Pushes reflection process                             & Utilizing the application pushes participants to reflect on data   protection                                                                                                   & 6                    \\ \hline

learning and exploration   process                    & Learning about privacy during exploration                                                                                                                                       & 11                   \\ \hline

improved understanding                                & Participants have learned something about data protection                                                                                                                       & 19                   \\ 
\hline

raised concern                                        & Participants show increased concern regarding data protection due to the   use of the tool                                                                                      & 17                   \\ \hline

improved awareness                                    & The use of the tool made particapants aware of privacy protection issues                                                                                                        & 46                   \\
\hline

\end{tabular}
\end{table}
\clearpage
\subsection{Questionnaire Results}
\label{sec:questionnaire}

\begin{figure}[!ht]
   \centering
   \small
   \includegraphics[width=0.9\linewidth]{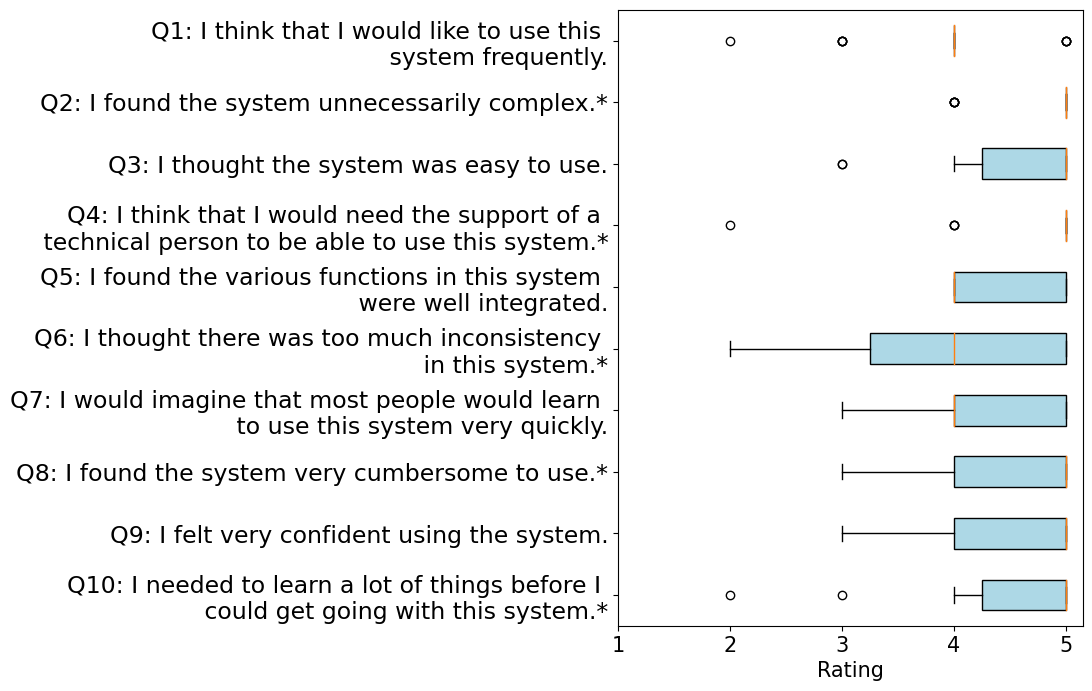}   
   \caption{System Usability Scale results (questions with * have inverted scores; higher values are always better)}
%    \Description{The plot is a boxplot visualization of the adjusted System Usability Scale (SUS) questionnaire results, where higher scores are always better. It displays ratings for 10 questions related to system usability, with the following observations:
% - Q1: The ratings range from about 2 to 5, with an outlier at rating 1.
% - Q2*: The ratings mostly span from 4 to 5, with a median of 4. Outlier below 3.
% - Q3: The ratings are between 3 and 5, with a median of 4. Outliers below 3.
% - Q4*: The ratings range from 2 to 5 with a median of 5.
% - Q5: The ratings span from about 4 to 5, with a median of 4.
% - Q6*: The ratings range from 2 to 5 with a median of 4.
% - Q7: The ratings fall between 3 and 5 with a median of 4.
% - Q8*: The ratings span from about 3 to 5, with a median of 5.
% - Q9: The ratings vary from 3 to 5 with a median of 5.
% - Q10*: The ratings range from about 2 to 5, with a median of 5 and
% an outlier at 2.
% The asterisks (*) next to certain questions denote that these questions were reverse-scored in the original SUS.}
   \label{fig:sus}
\end{figure}

\begin{figure}[!ht]
   \centering
   \includegraphics[width=0.9\linewidth]{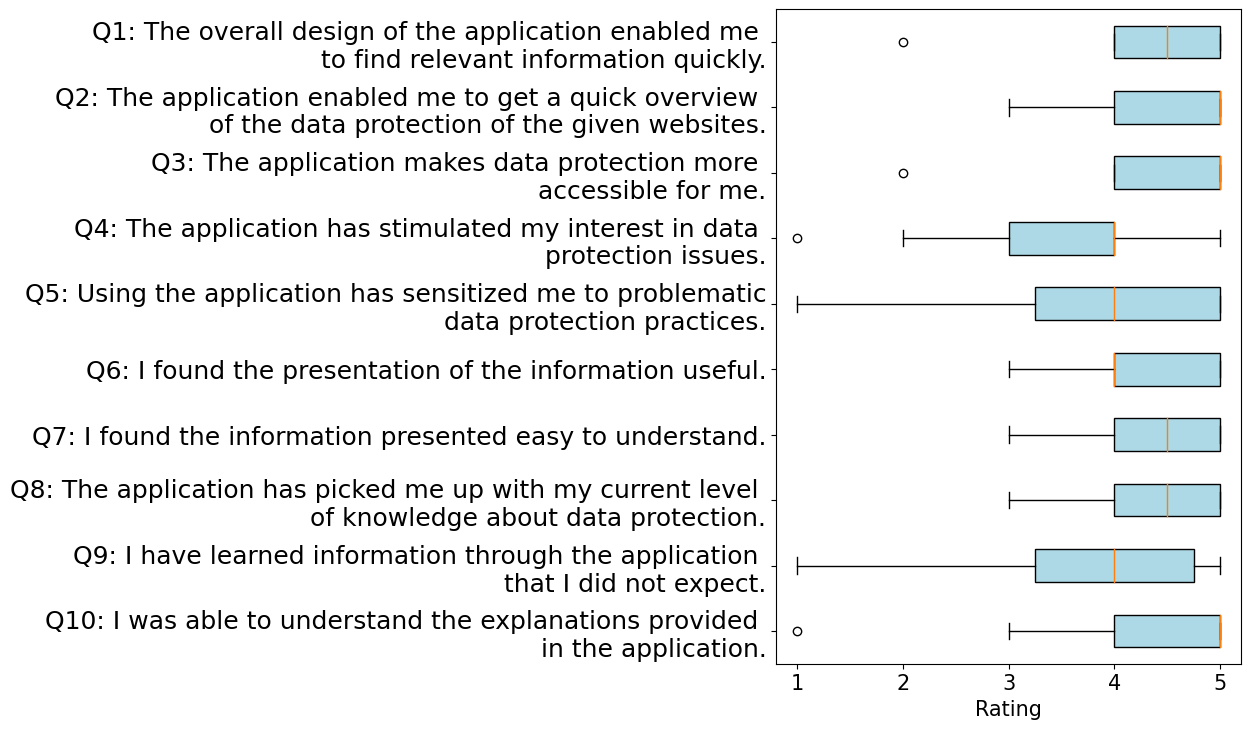}   
   \caption{Results of our questions on a 5-point Likert scale (1: strongly disagree; 5: strongly agree).}
%    \Description{The image displays a horizontal box plot that evaluates responses to a specific questionnaire. Each box plot corresponds to a different question in the questionnaire. The questions are listed on the y-axis, and the x-axis represents the rating on a scale of 1 to 5.
% - Each box plot contains a box, extending from the first quartile to the third quartile of the data, with a central line at the median.
% - Whiskers extend from the box to the minimum and maximum values within 1.5 times the interquartile range (IQR).
% - Outliers, data points that are outside the whiskers, are represented by circles.
% Observations:
% - Most questions have medians in the higher rating range (between 4 or 5).
% - Q1 and Q3 have the highest upper quartile close to the maximum rating.
% - The questions Q5 and Q9 have a wide distribution of responses.
% - There are some outliers present, especially for Q1 and Q10.
% }
   \label{fig:ownq}
\end{figure}

\end{document}